\documentclass[nofootinbib,twocolumn,prl,aps]{revtex4}
\usepackage{graphicx}

\begin{document}
\title{No black hole information puzzle in a relational universe}
\author{Rodolfo Gambini$^{1}$, Rafael A. Porto$^{2}$ and 
Jorge Pullin$^{3}$}
\affiliation {1. Instituto de F\'{\i}sica, Facultad de Ciencias,
Igu\'a 4225, esq. Mataojo, Montevideo, Uruguay. \\ 
2. Department of Physics, Carnegie Mellon University, Pittsburgh,
PA 15213\\
3. Department of Physics and Astronomy, Louisiana State University,
Baton Rouge, LA 70803-4001}
\date{December 15th 2003}

\begin{abstract}
The introduction of a relational time in quantum gravity naturally
implies that pure quantum states evolve into mixed quantum states. We
show, using a recently proposed concrete implementation, that the rate
at which pure states naturally evolve into mixed ones is faster than
that due to collapsing into a black hole that later evaporates. This
is rather remarkable since the fundamental mechanism for decoherence
is usually very weak. Therefore the ``black hole information puzzle''
is rendered de-facto unobservable.
\end{abstract}

\maketitle

Every physicist notes, upon first being introduced to quantum
mechanics, that the role of the variable ``t'' is somewhat
artificial. One is expected to believe the existence of a perfectly
classical external clock to the system in observation and to treat
time as a classical variable. This is 
clearly an approximation, that cannot ultimately be entirely 
accurate. After all, every clock will have some quantum fluctuations.
This is particularly true in situations where quantum gravity
effects may be of interest, since it is hard to imagine suitably
``external and classical'' clocks will be available at the relevant
energies.

How does one do quantum mechanics without classical clocks? The idea
consists in promoting all variables in the problem, including those
one may wish to choose as the clock variables, to quantum
operators. One then computes conditional probabilities for the
variables one wishes to observe assuming the variables chosen as clock
variables take certain desired values. If there is a single variable
that (at least for a while) behaves approximately classically, we can
call that variable $t$. Then the conditional probability of other
variables of interest taking a given value $x$ when $t$ takes a
certain value $t_T$, $P(x|t_T)$, will approximately satisfy a
Schr\"odinger equation.  When there is no
variable that is a good classical clock the relational approach is
still valid and the conditional probabilities exist, but the
interpretation of the probabilities as constituting a traditional
Schr\"odinger picture does not.

This framework appears remarkably natural to address the ``problem of
time'' in quantum gravity and was discussed in this context by Page
and Wootters \cite{PaWo}. Unfortunately, in the case of general
relativity, the presence of the constraints causes trouble. As
discussed in great detail by Kucha\v{r} \cite{Ku} it is not possible
to have a meaningful relational description in traditional canonical
quantum gravity. Since all observables of the theory are also
``perennials'' (they have vanishing Poisson brackets with the
Hamiltonian constraint) none of them is suitable as a ``clock''. Page
and Wootters attempt to bypass this by choosing to build the
relational framework in terms of quantities that do not have vanishing
Poisson brackets with the constraints (that is, they choose to work at
the kinematical level.) However, quantum mechanically, the states that
are annihilated by the constraints are expected to be distributional
within the space of kinematical states and do not lead to a good
probabilistic interpretation. Kucha\v{r} showed in model systems that
the resulting propagators are proportional to the Delta function and
therefore ``they don't propagate'' \cite{Ku}.

We have recently introduced new discretization scheme for general
relativity \cite{GaPu}. The resulting discrete theory contains
solutions that are arbitrarily close to those of general relativity,
yet is free of constraints. Therefore the major conceptual objections
to using the relational approach to solving the problem of time are
removed. We have discussed in some detail the application of this
approach in ref. \cite{GaPoPu}.

In quantum mechanics with a relational time, since the clock is not perfect,
it is inevitable that pure quantum states evolve into mixed 
quantum states. This is what brings us to the black hole information
paradox. In 1973 Hawking \cite{Ha} noted using quantum field theory on 
curved space-time techniques, that a black hole with mass $M$
will emit radiation as if it were at a temperature,
\begin{equation}
T={\hbar \over 8 \pi k M}
\end{equation}
where $k$ is Boltzmann's constant. Black holes are therefore not
ever-living anymore since they lose mass through the emitted
radiation. It is expected that eventually black holes evaporate
completely and the only thing left is the outgoing radiation.
This brings about the black hole information puzzle. Suppose one
prepares a pure quantum state and collapses it to form a  black hole.
Eventually all that is left is a thermal state of the outgoing
radiation, which is a mixed state. Therefore the pure state has
evolved into a mixed state.

In a world where unitarity rules, this is a problem. Unitarity rules
in the idealized world of Schr\"odinger quantum mechanics, with its
perfect clocks. In the more realistic  world of
quantum mechanics with a relational time, the lack of perfect clocks means
pure states evolve into mixed states naturally. The black hole
information ``problem'' will only arise if somehow black holes were
more efficient at destroying the coherence of pure states than
the lack of a perfect clock. Here we will attempt to estimate if
this is the case.

Several alternatives have been proposed as solutions of the black hole
information problem. A good brief summary can be found in the paper by
Giddings and Thorlacius \cite{GiTh}.  Hawking \cite{ha2} had proposed
that unitarity is lost in quantum mechanics due to interactions with
virtual black holes forming the ``space-time foam''. This approach has
been criticized on the grounds that it leads to the loss of the
conservation of energy \cite{hag}. It should be noted that our
proposal, although it has in common with Hawking's that it leads to
pure states evolving into mixed states, does conserve energy due to
the particular form of decoherence (it is a Lindblad \cite{lind} type
of evolution, but it is governed by the Hamiltonian and automatically
guarantees its conservation, see \cite{GaPoPu}.) It should be noted
that other effects, like the production of virtual pairs of black
holes \cite{hag} or the entanglement of the clock and the system upon
evolution could lead to lack of conservation of energy, but to a first
approximation energy is conserved in our approach. A second
alternative that was proposed as a solution to the paradox is that the
black hole does not evaporate completely, and a ``remnant'' containing
all the information is left. A challenge for proponents of this
approach is to find a satisfactory description of the remnants and to
avoid infinite production rates \cite{Gib}. Finally, a third avenue is
to attempt to find a way to send the information out with the outgoing
radiation in a process similar to quantum teleportation. A main
concern is to find dynamics that is non-local enough to achieve the
teleportation. Susskind has argued that string theory is non-local
enough in this sense \cite{suss}. A very recent and attractive
proposal along these lines is due to Horowitz and Maldacena
\cite{HoMa} in which they propose giving a boundary condition at the
singularity that transfers information to the outgoing radiation in a
process similar to quantum teleportation. Recently Gottesman and
Preskill have shown that one may still face non unitarities within
this scheme \cite{preskill}.

We now proceed to describe our proposal.  We have recently presented a
detailed calculation of the rate of decoherence of a pure state in the
relational picture emerging from discrete quantum gravity. The result
is that the time for losing complete coherence is given by
\cite{GaPoPu},
\begin{equation}
t_{\rm coherence\, loss} \sim {1 \over \omega^2 t_{\rm Planck}},
\end{equation}
where $\omega$ is the frequency associated with the spread of energy
of the possible states of the system under study. For a system with
two energy levels, it would be the transition frequency. The details
of the derivation of this formula are lengthy \cite{GaPoPu,GaPoPu2} and not
germane to what we are discussing, but it can be remarked that the
formula appears quite natural: a system with states with big energy
differences will tend to decohere more rapidly, and the natural scale
for decoherence due to quantum gravity ought to be the Planck scale.
So one could arrive to the preceding formula just on dimensional
analysis considerations. The resulting effect is very small. It is
conceivable that it could be observed experimentally in the future,
but it appears inaccessible to current technology \cite{GaPoPu}.

If we now consider a black hole, the energy fluctuations in the system
are characterized by, $kT={\hbar / 8 \pi M}$,  which
yields a frequency, $\omega = {k T / \hbar}={1 / 8 \pi M}$
(we use units in which $G=c=1$.) It is clear that extrapolating 
our formula, which was derived for a quantum mechanical system, to
a situation like a black hole can only be considered as a first 
approach, and to give a rough order-of-magnitude estimation of 
the effect. More detailed modeling is clearly needed, but at the
moment a reliable model with a detailed microscopic description of
the black hole seems beyond the reach of the state of
the art of the field.

Combining with our formula for the decoherence time we find 
that the system ``black hole plus pure state'' just due to the
possible quantum fluctuations of the black hole radiation, will
decohere in a time (from the point of view of an external 
observer at a fixed distance from the hole),
\begin{equation}
t_{\rm BH\, deco}\sim {(8\pi M)^2 \over t_{\rm Planck}}.
\end{equation}

On the other hand, the pure state will lose coherence through
the complete evaporation of the black hole in a time \cite{Kie},
\begin{equation}
t_{\rm evaporation} = (M/M_p)^3 t_{\rm Planck} 
\end{equation}

For the black hole evaporation to be faster than the natural
decoherence of the state, we have,
\begin{equation}
M<(8\pi)^2 M_{\rm Planck},
\end{equation}
that is, the black hole should be smaller than 631 Planck masses.
For such microscopic black holes the semiclassical picture in which
Hawking radiation is emitted is not valid and therefore one 
cannot formulate the information paradox in the traditional sense.
An analysis of this case would require a full quantum gravity
calculation.

The reader may wonder why is the mass limit so small? The reason is
that the loss of unitarity of the relational theory is produced by the
fluctuations in energy of the radiation emerging from the black hole,
and those fluctuations increase when the black hole is smaller. 
That is, a smaller black hole will evaporate faster, but it will
also decohere faster in the relational picture.
It should also be noted that our approach does not preclude the
loss of information through tunneling to another hypothetical 
semiclassical region beyond the singularity inside the black hole.
In fact it provides a natural mechanism to achieve such regions
\cite{GaPusmolin}.

Summarizing, we have argued that the correct way to view quantum
mechanics in a realistic universe where perfect classical clocks do
not exist is the relational approach. This approach has recently been
incorporated into the quantum gravity context via the consistent
discretization methods.  With the introduction of a relational time
pure states evolve into mixed states naturally, and we have here shown
that they do it fast enough that the black hole information puzzle
does not appear. It should be noted that in spite of this, the
rate of loss of coherence is really minute for everyday phenomena. For
instance the decoherence time for a Solar sized black hole is
$10^{41}$ years. However, the evaporation times are even larger, about
$10^{66}$ years.

To conclude, quantum mechanics with a relational time leads to a lack of coherence
that is minute and therefore does not interfere with most of the
physics we know, but it is large enough to eliminate the black hole
information puzzle.

This work was
supported by grant NSF-PHY0090091 and funds from the Horace Hearne
Jr. Institute for Theoretical Physics.


\begin{references}
\bibitem{PaWo} D.~N.~Page and W.~K.~Wootters, Phys.\ Rev.\ D {\bf
27}, 2885 (1983) 
\bibitem{Ku} K. Kucha\v{r}, ``Time and interpretations of quantum
gravity'', in ``Proceedings of the 4th Canadian conference on general
relativity and relativistic astrophysics'', G. Kunstatter, D.
Vincent, J. Williams (editors), World Scientific, Singapore (1992).
[Available at http://www.phys.lsu.edu/faculty/pullin/kvk.pdf]
\bibitem{GaPu} C. Di Bartolo, R. Gambini, J. Pullin,
Class. Quan. Grav. 19, 5475 (2002); R.~Gambini and J.~Pullin,
Phys. Rev. Lett. 90, 021301, (2003); R. Gambini, R.A. Porto and
J. Pullin, in ``Recent developments in gravity'', K. Kokkotas,
N. Stergioulas, World Scientific, Singapore, (2003)
[arXiv:gr-qc/0302064].
\bibitem{GaPoPu} R.~Gambini, R.~A.~Porto and J.~Pullin,
Class. Quan. Grav. {\bf 21} L51 (2004). 
\bibitem{Ha} S.~W.~Hawking, Commun.\ Math.\ Phys.\  
{\bf 43}, 199 (1975).
\bibitem{GiTh} S. Giddings, L. Thorlacius, in ``Particle and nuclear 
astrophysics and cosmology in the next millennium'', E. Kolb (editor), World
Scientific, Singapore (1996) [arXiv:astro-ph/9412046].
\bibitem{ha2}S.~W.~Hawking, Commun.\ Math.\ Phys.\  {
\bf 87}, 395 (1982).
\bibitem{hag} J. Ellis, J. Hagelin, D.V. Nanopoulos, and M. Srednicki,
Nucl. Phys. B241 (1984) 381; T. Banks, M.E. Peskin, and L. Susskind,
Nucl. Phys. B244 (1984) 125.
\bibitem{lind}G. Lindblad, Commun. Math. Phys. 48, 119 (1976)
\bibitem{Gib} G. Gibbons, in ``Fields and geometry'', ed. A.
Jadczyk (World Scientific, 1986); D. Garfinkle and
A. Strominger, Phys. Lett. B256 (1991) 146;
A. Strominger and S. Trivedi, Phys. Rev. {\bf D48}, 5778 (1993).
\bibitem{suss}L. Susskind, Phys. Rev. Lett. 71 (1993) 2367; Phys.
Rev. D49 (1994) 6606.
\bibitem{HoMa} G.~T.~Horowitz and J.~Maldacena, JHEP {\bf 0402} 008
(2004).  
\bibitem{preskill} 
D.~Gottesman and J.~Preskill, JHEP {\bf 0403} 026 (2004). 
\bibitem{GaPoPu2} R. Gambini, R. Porto, J. Pullin, 
New. J. Phys. {\bf 6}, 45 (2004).
\bibitem{Kie} C.~Kiefer, in ``Decoherence and entropy in complex
systems'', H.-T. Elze (editor), Springer-Verlag, New York (2003)
[arXiv:gr-qc/0304102].
\bibitem{GaPusmolin}R. Gambini, J. Pullin,  
Int. J. Mod. Phys. {\bf D12} 1775 (2003). 
\end{references}
\end{document}